\documentstyle[prb,psfig,epsfig,aps]{revtex}

\input{epsf}

\begin{document}
\draft
\title {Quantum resonances in a single plaquette of Josephson junctions:\\ 
excitations of Rabi oscillations}

\author{M. V. Fistul}
\address{Max-Planck Institut f\"ur Physik Komplexer Systeme,
D-01187, Dresden, Germany}
\date{\today}

\wideabs{ %REVTeX 3.1 feature

\maketitle

\begin{abstract} We present a theoretical study of a quantum regime of 
the {\it resistive} (whirling) state of dc driven
anisotropic single plaquette containing three small Josephson junctions.
The current-voltage characteristics of such a system display
resonant steps that are due to the resonant 
interaction between the time dependent Josephson current 
and the excited electromagnetic oscillations (EOs). 
The voltage positions of the resonances are determined by the quantum interband transitions of 
EOs. We show that in the quantum regime 
as the system is driven on the resonance, coherent Rabi oscillations  
between the quantum levels of EOs occur. At variance with the classical regime 
the magnitude and the width of resonances are determined by the frequency 
of Rabi oscillations that in turn, depends in a peculiar manner on
an externally applied magnetic field and the parameters of the system.
\end{abstract}

}

%end of WideAbs
%\newpage

Various quantum effects predicted and observed in {\it macroscopic} systems 
\cite{QLegg,Tinkham,Klarke,Moya,Moya1,Lukens,SchonHanIngold}
have attracted a great attention as it allows to understand the foundation 
of quantum mechanics \cite{QLegg} and the applicability of quantum mechanics 
to dissipative systems.\cite{SchonHanIngold,Dekker,QLegg1}  
Moreover, the interest to this field has been boosted by the possibility to use 
macroscopic quantum coherent effects for quantum computation.
\cite{Moya,Lukens,Walraff} 

In this field of study Josephson coupled systems consisting of a few  
interacting Josephson junctions, are of special interest. These systems 
contain a large number of particles and still, their behaviour is determined 
by the macroscopic variables, namely Josephson phases $\varphi_i (t)$. Moreover,
the dynamics of Josephson phases can be controlled by an externally 
applied magnetic field $H_{ext}$ and dc bias $\gamma$, and
at low temperatures the number of quasi-particles is extremely small and  
therefore, the dissipation caused by the quasi-particle current is also small.  

Indeed, the peculiar macroscopic quantum effects such as 
tunneling and resonant tunneling of Josephson phase, discrete energy levels 
have been observed in single Josephson junctions, SQUID systems etc.
In the presence of an
{\it externally} applied microwave radiation 
the enhancement of both tunneling \cite{Klarke} 
and resonant tunneling of Josephson phase, have been also observed.
\cite{Moya,Lukens} 
These effects can be considered as an evidence of the  
quantum coherent dynamics, i. e. the presence of coherent Rabi oscillations 
in macroscopic systems.
The majority of macroscopic quantum effects have been studied
as the Josephson junctions were biased in the superconducting state, i. e. 
zero dc voltage state, and 
the quantum mechanical behaviour of the resistive state of Josephson 
coupled systems has not been analyzed.

It is also well known 
that in the classical regime the resistive state of Josephson coupled systems 
displays intrinsic Josephson current oscillations.
The oscillating Josephson current can excite 
the electromagnetic oscillations (EOs) in the superconducting loops, 
and in turn, these EOs resonantly  
interact with the time dependent Josephson current.
In the case of a weak damping such an interaction leads to a pronounced 
resonant step in the current- voltage
characteristics ($I$-$V$ curves). The voltage position of the resonance 
$V$ is determined by the characteristic frequency of EOs $\omega_0$ as
\begin{equation} \label{ResVolt:Cl}
V~=~\frac{\hbar \omega_0}{2e}~~.
\end{equation}
The magnitude of the resonance depends on an externally applied 
magnetic field $H_{ext}$ and the width of the resonance 
is determined by the damping parameter  
that, in the classical regime, is due to the presence of a quasiparticle 
(dissipative) current. 

In this paper I present a theoretical (semiclassical) 
analysis of the quantum 
coherent effects in the {\it resistive} (whirling) state of  
a dc driven single anisotropic plaquette containing three small 
Josephson junctions.  
This system consists of two {\it vertical} junctions parallel to the
bias current $\gamma$ and a {\it horizontal} junction in the transverse direction, 
as presented in Fig. 1.

The dynamics of the system crucially depends on two parameters:
the anisotropy $\eta=\frac{I_{cH}}{I_{cV}}$, where 
$I_{cH}$ and $I_{cV}$ are respectively the critical currents of horizontal 
and vertical junctions, and the discreteness parameter 
(normalized inductance of the cell), $\beta_L$.
Moreover,
the quantum effects are 
enhanced in the limit of a small Josephson energy $E_J~=~
\frac{\hbar I_{cV}}{2e}~\leq~\hbar \omega_p$, where $\omega_p$ is 
the plasma frequency. In this limit it is also naturally to apply an external 
charge $Q$ to the horizontal junction (see, Fig. 1). 
This charge controls the frequency of transitions between different 
quantum levels of EOs. 
Such a system presents a simplest case allowing to couple the Josephson current oscillations 
with a  nonlinear oscillator (horizontal junction), and therefore, 
to remove the quantum-classical 
correspondence of a harmonic oscillator\cite{QLegg,Klarke} and to 
observe the quantum effects in the resistive state.
Note here, that the coherent quantum-mechanical behaviour of 
a single plaquette of Josephson junctions
biased in the superconducting state, have been studied in details 
in Refs.\cite{Moya,Moya1}

 \begin{figure}
  \centerline{\epsfig{file=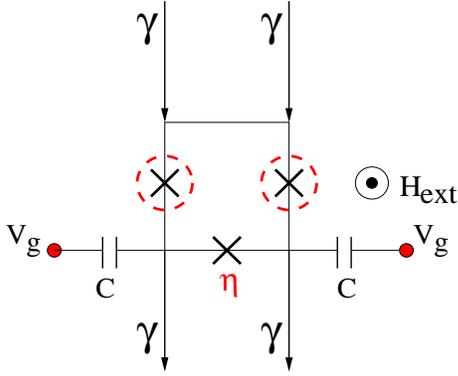,height=5cm,width=6cm}}
 \vspace{2mm}
  \caption{Sketch of the plaquette with three Josephson junctions 
  (marked by crosses). Arrows indicate the directions of external current flow,  
  (dc bias $\gamma$). A gate voltage $V_g$ allows to introduce a charge 
  $Q$ in the system, and the gate capacitors are, $C=\alpha C_H$. 
  The externally applied magnetic field $H_{ext}$ is also shown. 
  Dashed circles denote junctions in the resistive (whirling) state.}
   \label{Fig:exp}
\end{figure}

We found that the quantum 
effects  affect the resonant interaction between EOs and 
oscillating Josephson current, and  the voltage positions of the 
resonant steps are determined by the discrete {\it 
energy levels} of a nonlinear oscillator.\cite{comment} Moreover, the obtained   
peculiar dependence of the magnitude and the width of resonances 
on the externally applied magnetic field $H_{ext}$ 
can be considered as fingerprints of {\it coherent Rabi oscillations} excited 
as the quantum transitions in the spectrum of EOs occur.  

The dynamics of a single plaquette of three Josephson junctions is determined 
by time dependent Josephson 
phases of vertical junctions $\varphi^v_{1,2}(t)$, and the horizontal 
junction $\varphi_h (t)$. The dynamics of the Josephson phases is described by 
the Lagrangian
$$
L~=~E_J\{\frac{1}{2{\omega_p}^2}[(\dot \varphi^v_1)^2 + 
(\dot \varphi^v_2)^2+
\eta(\dot \varphi_h-\alpha v_g)^2] +\cos (\varphi^v_1) + 
$$
$$
+\cos (\varphi^v_2) +\eta \cos (\varphi_h) +\gamma (\varphi^v_1+\varphi^v_2) -
$$
\begin{equation} \label{Lagr}
-\frac{1}{\beta_L}(\varphi^v_1-\varphi^v_2 +\varphi_h+2\pi f)^2\}
\end{equation}
Here, 
the dc bias $\gamma$  
is normalized to the critical current of the junction $I_{cV}$, and 
the normalized gate voltage $v_g~=~2\pi V_g/\Phi_0$.
The externally applied magnetic field $H_{ext}$ is characterized by 
the frustration $f=\frac{\Phi_{ext}}{\Phi_0}$, i. e. 
the magnetic flux threading the cell normalized to the magnetic flux quantum.  

In the case of a low-inductance environment 
both quantum and temperature fluctuations weakly alter 
the Josephson phases of the vertical junctions that are in the whirling 
state.\cite{IngNaz} Thus, the Josephson phases can be naturally decomposed 
$$
\varphi^v_{1}(t)~=~\omega t -\pi f -\xi(t)~~,
$$
\begin{equation} \label{Sol}
\varphi^v_{2}(t)~=~\omega t +\pi f +\xi(t)~~.
\end{equation}
The frequency $\omega~=~2eV/\hbar$ is determined by the dc voltage $V$ across 
the junction. 
As a result we find that  
the supercurrent flowing through the vertical junctions $I_s$ 
is expressed in the form:
\begin{equation} \label{SuperCurr}
I_s~=~I_{cV}<\sin (\omega t) \cos( \pi f +\xi(t))>~~,
\end{equation}
where $<...>$ means the time-average procedure. 

Next, to simplify the analysis, we consider a small plaquette of 
Josephson junctions as the discreteness parameter $\beta_L~<<~1$. In this case
the relationship $\xi(t)~=~\varphi_h/2$ is valid, and the system is 
characterized by  one degree of freedom $\xi$.
Introducing the canonical momentum $p_\xi~=~\partial L/\partial \dot \xi$ and 
the corresponding  operator of momentum 
$\hat p_\xi~=~-i\hbar \partial/\partial \xi$,\cite{Moya} we arrive at
the time-dependent Hamiltonian 
$$
\hat H (t)~=\hat H_0-2E_J
\cos (\pi f+\xi) 
cos (\omega t)~~,
$$
\begin{equation} \label{Hamilton}
\hat H_0~=~ ~\frac{\omega_p^2}{E_J (4+8\eta)}(\hat p_{\xi}-4\eta \alpha v_g)^2-
E_J\eta \cos 2\xi~~.
\end{equation}   
Here, $\hat H_0$ is  
the Hamiltonian of the autonomous nonlinear oscillator, where the first term 
presents the total charging energy of the system and the second term is the 
Josephson energy of the horizontal junction.
The last term in $\hat H(t)$ presents an {\it intrinsic} magnetic field 
dependent coupling between the time dependent Josephson current and EOs.  

We are interested in the resonant interaction between the ac Josephson current 
and EOs, and thus, two relevant energy 
levels $E_m$ and $E_n$ of the Hamiltonian $\hat H_0$, namely
$ \omega_{nm}(v_g)~=~E_n(v_g)-E_m(v_g)~\simeq~\hbar \omega$, are important 
for our problem. These energy levels may be controlled by an 
externally applied gate voltage $v_g$. 
Because a nonlinear oscillator 
has no coinciding frequency differences $\omega_{nm}$  we may
truncate our system to the two-level system. 
With this crucial assumption the Hamiltonian $\hat H(t)$ is written in a  
simple form:
$$
\hat H (t)~=\frac{\omega_{nm}}{2} \hat \sigma_z+E_J
(a_{nn}-a_{mm})\cos(\omega t) \hat \sigma_z
$$
\begin{equation} \label{Hamilt:TwoLevel}
-2E_J
|a_{nm}|\cos(\omega t) \hat \sigma_x~~,
\end{equation} 
where the matrix elements $a_{nm}$ are
\begin{equation} \label{MatrElem}
a_{nm}~=~\int_0^{2\pi} d\xi \psi_n^\star (\xi; v_g) \psi_m(\xi; v_g)cos(\pi f+\xi) ~~.
\end{equation} 
Here, $\psi_{n,m}(x;v_g)$ are the gate- voltage dependent 
wave functions of the 
autonomous nonlinear oscillator, 
and $\hat \sigma_{x,z}$ are the Pauli matrices. 
 
Next, we use the standard density matrix approach.\cite{Blum}
In the case of a weak damping the corresponding time-dependent equation for the 
density matrix $\hat \rho (t) $ is taken in the form
\begin{equation} \label{EqDensMatr}
 \hbar  \dot {\hat \rho} (t) ~=~-i[{\hat H ( t)}, \hat \rho (t) ] +
[\hat H_R, (\hat {\rho}(t)-\rho_\beta)] ~~,
\end{equation} 
where $\rho_\beta$ is the equilibrium density matrix, and the dissipative 
operator $\hat H_R$ characterizes the various relaxation processes. 
In a simplest case this operator is described by two damping  
parameters $\nu_{1,2}$.\cite{Blum,BlochEq}
By making use of (\ref{SuperCurr}) the supercurrent $I_s$ is 
expressed through the quantum-mechanical average of of the operators 
$\hat \sigma_{x,z}$ as
$$
I_s~=~I_{cV}<|a_{nm}|sin(\omega t) \bar 
Tr\{\hat \rho(t) {\hat \sigma_x} \}+
$$
\begin{equation} \label{SuperCurr2}
+\frac{a_{nn}-a_{mm}}{2}sin(\omega t) \bar 
Tr\{\hat \rho(t) {\hat \sigma_z} \}>~~.
\end{equation}
Eq. (\ref{EqDensMatr}) is a particular case of the well-known Bloch 
equations\cite{Blum,BlochEq}, and by using the rotation wave 
approximation\cite{BlochEq} we finally obtain 
$I_s$ as
\begin{equation} \label{SuperCurr3}
I_s~=~\frac{2eE_J^2}{\hbar^2} |a_{nm}|^2 \frac{\nu_2}{(\omega-\omega_{nm})^2+
 \nu_2^2+2(\frac{E_J}{\hbar})^2(\frac{\nu_2}{\nu_1})|a_{nm}|^2}
\end{equation}

Thus, Eq. (\ref{SuperCurr3}) shows that the current-voltage characteristics 
of the plaquette with three small Josephson junctions can display a number of 
resonances. The physical origin of these resonances is the resonant 
absorption of the ac Josephson oscillations by the horizontal Josephson 
junction being in the superconducting state. 
The voltage positions of the resonances $V~\simeq~ \hbar\omega_{nm}/2e$ are 
mapped to the various transitions occurring in the spectrum of EOs 
(the Josephson phase of the horizontal junction).
The
width of the resonances is determined by the relaxation processes as 
$\nu~\gg ~E_J|a_{nm}|/\hbar$, or the frequency 
$\omega_R~\simeq~\frac{E_J|a_{nm}|}{\hbar}$ of coherent Rabi oscillations in the opposite 
limit ($\nu~\ll ~E_J|a_{nm}|/\hbar$).
The maximum value of the magnitude of the resonance depends on the 
damping parameters $\nu_{1,2}$, and may reach the value 
$I_s^{max}~\simeq~ e\nu_2$. Note here that we  
assumed the low temperature regime ($T~\leq~\hbar \omega_{nm}$) and 
did not take into account processes involving
multi-photon interactions between the ac Josephson current and EOs. 
These multi-photon interactions lead  
to additional subharmonic resonances 
($\omega~\simeq~\omega_{nm}/k$) with smaller magnitude. 

The spectrum $E_n(v_g)$ and the 
corresponding wave functions are found as periodic solutions of the Schr\"odinger 
equation:
\begin{equation} \label{SchrEq}
 \hat H_0 \psi_n(\xi;v_g)~=~E_n(v_g)\psi_n(\xi;v_g)~.
\end{equation}
It is well known that the spectrum $E_n(v_g)$ of Eq. (\ref{SchrEq}) contains an infinite number of 
bands and is controlled by the gate-voltage $v_g$.\cite{LZ,Schon}
Although  
the solutions of (\ref{SchrEq}) can be analized
by making use of the Mathiew functions \cite{Abram}
for arbitrary ratio
$E_J/\hbar \omega_p$, here we consider the regime of small Josephson energy, 
$E_J~\ll~\hbar \omega_p$, where all results are simplified. 
In this limit, and at low temperatures as the transitions between 
the ground state and the excited levels are important, we obtain 
\begin{equation} \label{Spectrum}
\omega_{n0}~\simeq~\frac{\omega_p^2}{E_J (4+8\eta)}|n(n-8\eta \alpha v_g)|
\end{equation}
The typical dependence of $E_n(v_g)$ and a number of transitions are shown in 
Fig. 2a.
\begin{figure}
  \centerline{\epsfig{file=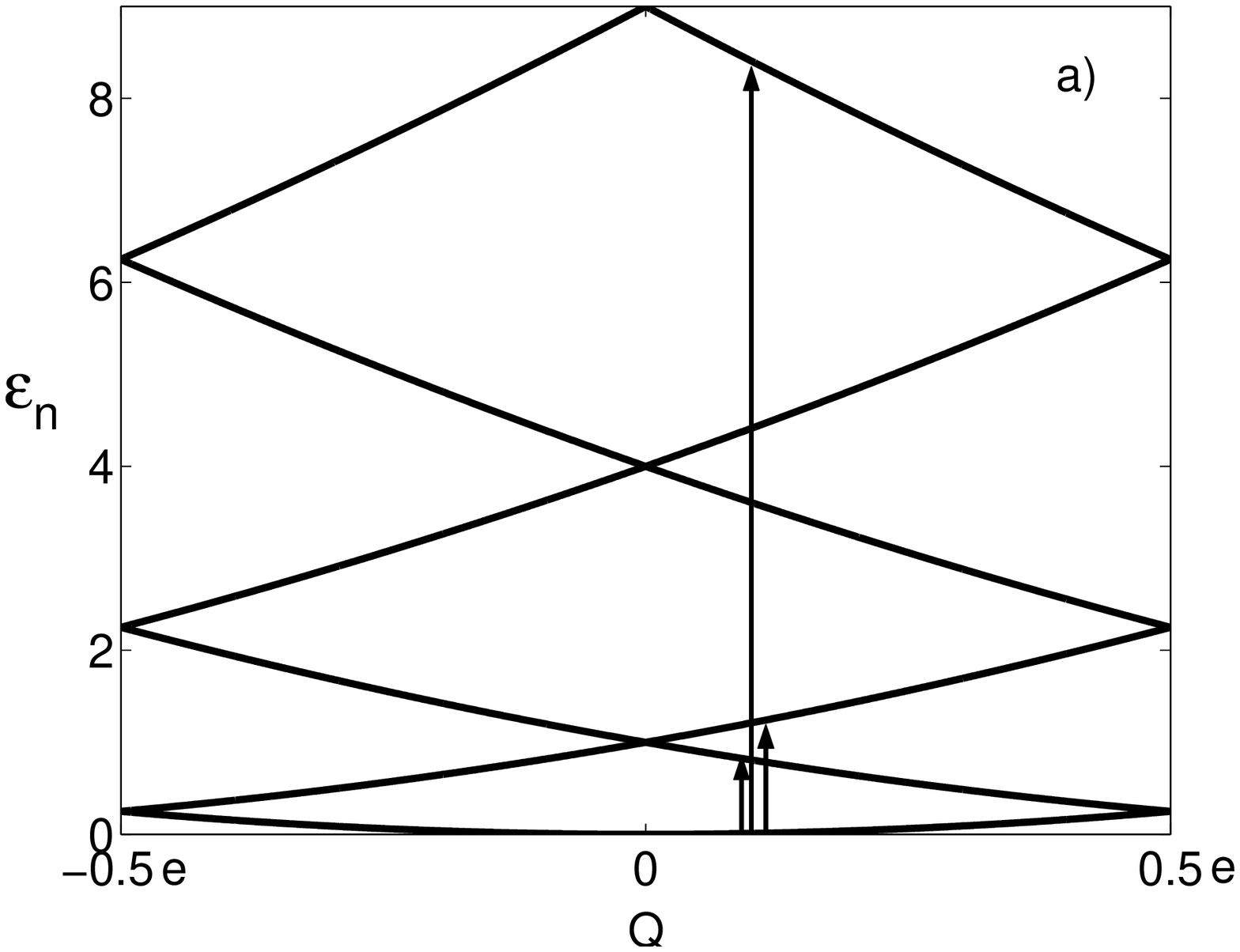,height=5cm,width=6cm}}
  \centerline{\epsfig{file=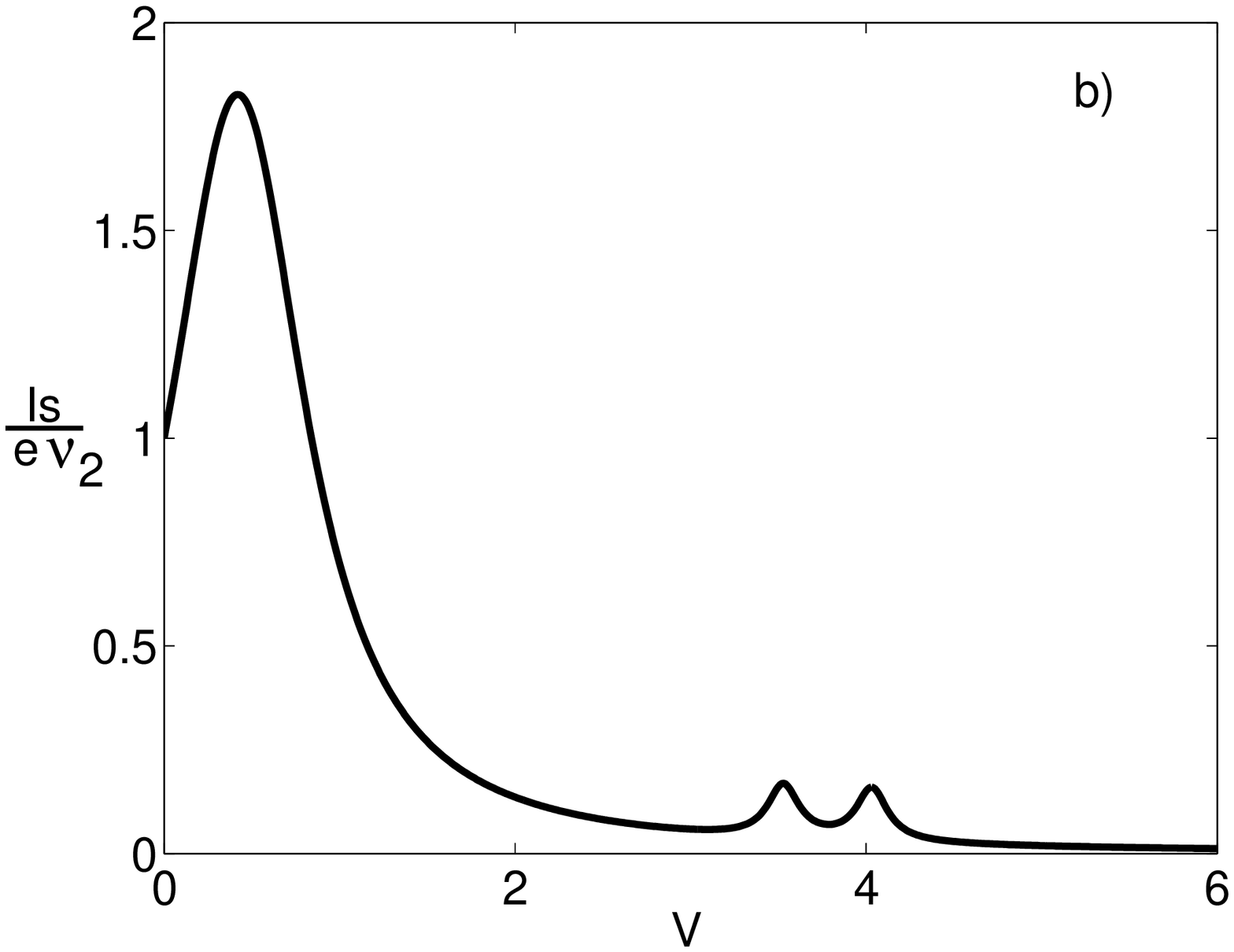,height=5cm,width=6cm}}
 \vspace{2mm}
  \caption{a) The dependence of 
  $\epsilon_n(v_g)~=~\frac{E_J (4+8\eta) E_n(v_g)}{\hbar^2 \omega_p^2}$ 
  in the limit of zero Josephson energy. The induced charge $Q~=~4e\nu \alpha v_g$. The arrows show possible 
  transitions in the spectrum of EOs.
  b) The quantum resonances in the current-voltage characteristics. 
  The resonances correspond to the transitions presented in Fig. 2a. The  
  voltage is normalized to the plasma frequency $\hbar \omega_p/2e$.
  The parameters are $\eta~=~1$, $\nu_1~=~\nu_2~=~0.1\omega_p$, 
  $\frac{E_J}{\hbar\omega_p}~=~0.3$, $Q=0.1$.
  }
   \label{Fig:2}
\end{figure}
 
By making use of a perturbation theory the relevant matrix elements 
$a_{nm}$ are obtained in this limit. Thus, e.g.
$a_{\pm 1~0}~\simeq~1$, 
$a_{\pm 3~ 0}~\simeq~(\frac{E_J}{\hbar \omega_p})^2$. The transition 
$0 \rightarrow 2$ is not appearing in the $I$-$V$ curve because the
matrix element $a_{2~0}$ is rather small. It is due to a specific symmetry of
the potential energy ($\propto \cos 2\xi$) in the Hamiltonian $\hat H_0$. 
The calculated resonant 
current-voltage characteristics is presented in Fig. 2b.
As the Josephson energy $E_J$ is small, the voltage positions 
of the resonances are strongly affected by the gate voltage $v_g$ but the 
widths and the magnitudes of the resonances weakly depend on the 
externally applied magnetic field $H_{ext}$. In the opposite case as the  
Josephson energy is large, $E_J~\gg~\hbar \omega_p$, the situation is reversed:
the voltage positions of the resonances are weakly altered by $v_g$ but 
the width 
of the resonances strongly depends on $H_{ext}$.
In the case of intermediate values of $E_J~\simeq~\hbar \omega_p$ 
the strong dependence of the resonant current-voltage characteristics on  
both parameters $v_g$ and $H_{ext}$ is found.

In conclusion we have shown that a particular system of 
a single plaquette containing three small Josephson junctions display 
resonances in the $I$-$V$ curve. These resonances are due to the 
resonant absorption of intrinsic ac Josephson oscillations and are the 
fingerprints of various transitions between the discrete energy levels of 
the macroscopic Josephson phase. The coherent quantum-mechanical 
dynamics of these transitions may be controlled by variation of the bias
current $\gamma$, the gate voltage $V_g$ and an externally 
applied magnetic field $H_{ext}$.  
Finally, we note that similar quantum resonances may be found also 
in more complex Josephson (or mixed) coupled systems, e.g. the 
inductively coupled dc and RF SQUIDs, dc SQUID and quantum dots \cite{Ustinov},
etc.\cite{comment} The measurements of these resonances and their dependence 
on the parameters of the system should allow to study in detail
the coherent quantum-mechanical dynamics of macroscopic variables.

I thank S.-G. Chung, S. Flach, P. Hakonen, and A. V. Ustinov  
for useful discussions. 
%%%%%%%%%%%%%%%%%%%%NEW%%%%%%%%%%%%%%%%%%%%%%%%%%%%%%%%%%%%%%%%%%%%%%%
%%%%%%%%%%%%%%%%%%%%%%%%END NEW%%%%%%%%%%%%%%%%%%%%%%%%%%%%%%%%%%%

\end{document}